\documentclass[12pt,a4paper]{article}
\usepackage{amssymb,amsfonts,amsmath}

\title{Quantum(-like) formalization of common knowledge: 
Binmore-Brandenburger operator approach}

\author{Irina Basieva and Andrei Khrennikov
\\
International Center for Mathematical Modeling\\
 in Physics and Cognitive Sciences \\
 Linnaeus University,  V\"axj\"o-Kalmar, Sweden\\
\\Prokhorov General Physics Institute \\
 Russian Academy of Science, Moscow, Russia}

\begin{document}

\maketitle              

\begin{abstract} We present the detailed account of the quantum(-like) viewpoint to  
common knowledge. The Binmore-Brandenburger operator approach to the notion of common knowledge 
is extended to the quantum case. We develop a special quantum(-like) model of common knowledge 
based on information representations of agents which can be operationally represented 
by Hermitian operators. For simplicity, we assume that each agent constructs her/his 
information representation by using just one operator. However, different 
agents use in general representations based on noncommuting operators, i.e.,
incompatible representations.  The quantum analog of basic system of common knowledge features  
${\cal K}1-{\cal K}5$ is derived. 
  
{\bf keywords:} common knowledge, Binmore-Brandenburger operator approach, quantum(-like) decision making
\end{abstract}

\section{Introduction}

Common knowledge plays the crucial role in establishing of social conventions (as was firstly pointed out at the scientific level by 
David Hume in 1740). And the last 50 years were characterized by development of numerous formal (sometimes mathematical, but sometimes 
not) models of common knowledge and operating with it. One of the most useful mathematical formalizations is due to 
Binmore-Brandenburger   \cite{Brandenburger}. Starting with classical measure-theoretic model of 
probability theory (Kolmogorov, 1933) they elaborated the formal  approaches to the notion of common knowledge.
The operator approach  Binmore-Brandenburger is based on the notion of agents' knowledge operators 
$K_i.$ 

Common knowledge models play an important role in decision making theory, game theory, and cognitive psychology leading, in particular,
to the Aumann theorem on the impossibility to agree on disagree in the presence of nontrivial common knowledge and the common prior \cite{Aumann}, 
\cite{Aumann1}.
Recently the quantum(-like) decision theory flourished as the result of  
the fruitful cooperation of the psychological and quantum probability communities, see, e.g., the monographs \cite{40}-\cite{33}. 
Therefore it is a good time to present quantum(-like) formalization of the notion of 
common knowledge and to extend Aumann's argument on ``(dis)agree on disagree'' to the quantum case. The latter is discussed 
in another paper of the authors presented to QI2014 \cite{QI2014} (see also this paper for extended bibliography 
on quantum cognition). And in this note we present the detailed account of the quantum(-like)
approach to common knowledge. We start with a quantum analog of Aumann's definition of knowing of an event $E$ for the fixed 
state of the world $\omega \in  \Omega.$ Then we introduce the knowledge operator corresponding to such a  notion of 
knowing. We show that this quantum (super)operator satisfies the system of axioms ${\cal K}1-{\cal K}2$ for  
the Binmore-Brandenburger   \cite{Brandenburger} knowledge operators. Thus the quantum knowledge 
operator can be considered as a natural generalization of the classical knowledge operator. One of possible interpretations of 
such generalization is that the collection of possible information representations of the world by agents is extended. Such nonclassical
information representations are mathematically given by spectral families of Hermitian operators (``questions about the world'' stated
by the agents). In this operator framework we introduce hierarchically defined common knowledge (which was used to 
formulate the quantum(-like) analog of the (anti-)Aumann theorem \cite{QI2014}).  

In classical theory the operator definition of common knowledge  matches with the heuristic    
viewpoint on common knowledge; for two agents $i=1,2$, 

\medskip

${\bf COM_KN}$ An event $E$ is common knowledge at the state of the world
$\omega$ if 1 knows $E,$ 2 knows $E,$ 1 knows 2 knows $E,$ 2 knows 1 knows $E,$ and so on... 

\medskip

Our quantum(-like) notion of operator 
common knowledge matches with human intuition as well. (The difference is mathematical formalization of knowing.)

To simplify mathematics, we proceed with {\it finite dimensional state spaces.}
Generalization to the infinite dimensional case is evident, but it will be based on more advanced mathematics.

We also remark that our model of quantum(-like) formalization of common knowledge can be generalized by using the formalism 
of open quantum systems leading to questions represented by positive 
operator valued measures, cf. \cite{44}, \cite{POVM0}, \cite{POVM}, or even more general operator valued measures \cite{IB}. 
(In principle, there is no reason to expect that the operational description of cognitive phenomena, psychology, and economics would be 
based on the exactly the same mathematical formalism as the operational description of physical phenomena. Therefore we cannot exclude 
that some generalizations will be involved, see again \cite{IB}. )
However, at the very beginning we would like to separate the mathematical difficulties from the 
formalism by itself; therefore we proceed with quantum observables of the Dirac-von Neumann class, Hermitian operators and projector valued 
operator measures.

\section{Set-theoretic model of common knowledge}

In the classical set-theoretic model  events (propositions) are represented by subsets of some set $\Omega.$ Elements of this set 
represent all possible states of the world (or at least states possible for some context).
In some applications, e.g., in sociology and economics, $\Omega$ represents possible states of affairs. 
Typically considerations are reduced to finite (or countable) state spaces. In the general case, one has to proceed
as it common in classical (Kolmogorov) model of probability theory and consider a fixed $\sigma$-algebra of subsets 
of $\Omega,$ say ${\cal F},$ representing events (propositions).   

There is a group of agents (which are individual or collective cognitive entities); typically the 
 number of agents is finite, call them $i=1,2,..., N.$ These individuals are about to learn the answers to 
various multi-choice {\it questions} about the world (about the state of affairs),   
to make observations. In the Bayesian model agents assign  prior probability distributions for 
the possible states of the world; in many fundamental considerations such as, e.g., Aumann's theorem, 
it is assumed that the agents set the common prior distribution $p,$ see \cite{QI2014} for more details.
Here one operates with the classical Kolmogorov probability space $(\Omega, {\cal F}, p).$ In this note
we shall not study the problem of the prior update, see again     \cite{QI2014}. Therefore at the classical level 
our considerations are restricted to set-theoretic operations. 

Each agent creates its information representation for possible states of the world based on its own
possibilities to perform measurements, ``to ask questions to the world.'' Mathematically these representations 
are given by partitions of $\Omega: {\cal P}^{(i)}= (P_j^{(i)}),$ where $\cup_j P_j^{(i)}= \Omega$ and 
$P_j^{(i)} \cap P_k^{(i)} \emptyset, j\not=k.$ Thus an agent cannot get to know the state of the world $\omega$ 
precisely; she can only get to know to which element of its information partition $P_j^{(i)}= P^{(i)}(\omega)$ 
this $\omega$ belongs. The agent $i$ knows  an event $E$ in the state of the world $\omega$ if 
\begin{equation}
\label{KNOW}
P^{(i)}(\omega) \subset E.
\end{equation}
Let $K_i(E)$ be the event ``$i$th agent knows $E$'':
\begin{equation}
\label{KNOW7}
K_i E =\{\omega \in \Omega: P^{(i)}(\omega) \subset E \}.
\end{equation}
As was shown by Binmore-Brandenburger \cite{Brandenburger}, the {\it knowledge operator} $K_i$ has the following properties:
\[{\cal K}1: \;\;\;K_i E \subset E\]
\[{\cal K}2: \;\;\; \Omega \subset K_i \Omega\]
\[{\cal K}3: \;\;\; K_i (E\cap F) = K_i E \cap K_i F\]
\[{\cal K}4:  \;\;\; K_i E \leq  K_i K_i E\]
\[{\cal K}5: \;\;\; \overline{K_i E}  \leq K_i \overline{K_i E}\]
Here, for an event $E,$ $\bar{E}$ denotes its complement.  
We remark that one can proceed another way around \cite{Brandenburger}: to start with ${\cal K}1-{\cal K}5$ as the system 
of axioms determining the operator of knowledge and then derive that such an operator has the form (\ref{KNOW7}).

The statement ${\cal K}1$ has the following meaning:
 if the $i$th agent knows $E$, then $E$ must be the case; the statement 
${\cal K}2:$ the $i$th agent knows that some possible state of the world in $\Omega$ occurs;
${\cal K}3:$ the $i$th agent  knows a conjunction if, and only if, i knows each conjunct;
${\cal K}4:$  the $i$th agent knows $E$, then she knows that she knows $E;$ 
${\cal K}5:$  if the agent does not know an event, then she knows that she does not know.

\section{Quantum(-like) scheme}

Let $H$ be (finite dimensional) complex Hilbert space;
denote the scalar product in $H$ as $\langle\cdot \vert \cdot \rangle.$ 
For an orthogonal projector $P,$ 
we set $H_P= P(H),$ its image, and vice versa, for subspace $L$ of $H,$
the corresponding orthogonal projector is denoted by the symbol $P_L.$

In our  model the ``{\it states of the world}'' are given by pure states (vectors of norm one); events 
(propositions) are represented by orthogonal projectors. 
As is well known, these projectors form a lattice (``quantum logic'') with the operations 
corresponding to operations on orthocomplemented subspace lattice of complex Hilbert space $H$
(each projector  $P$ is identified with its image-subspace of $H_P).$

Questions posed by agents are mathematically described by self-adjoint operators, say $A^{(i)}.$   
Since we proceed with finite-dimensional state spaces, 
$
A^{(i)} = \sum_j a_j^{(i)}   P_j^{(i)},
$
where $(a_j^{(i)})$ are real numbers, all different eigenvalues of $A^{(i)},$ and $(P_j^{(i)})$  are the orthogonal projectors onto
the corresponding eigen-subspaces.  Here $(a_j)$  encode possible answers to the 
question of the $i$th agent. The system of projectors ${\cal P}^{(i)} = (P_j^{(i)})$ is 
the spectral family of $A^{(i)}.$ Hence, for any agent $i,$ it is   
a ``disjoint partition of unity'':
$\vee_{k} P_k^{(i)}  =I, \; P_k^{(i)}\wedge P_m^{(i)} =0, k\not=m,$ or equivalently 
$
\sum_{k} P_k^{(i)}  =I, \; P_k^{(i)} P_m^{(i)} =0, k\not=m. 
$
This spectral family can be considered as 
information representation of the world by the $i$th agent.
In particular, ``getting the answer $a_j^{(i)}$'' is the event which is mathematically described 
by the projector $P_j^{(i)}.$

If {\it the state of the world}\footnote{The general discussion on the meaning of the state of the world is presented in our second conference
paper \cite{QI2014}. It is important to remark that in models of qunatum cognition states are typically not physical states, but information 
states. They give the {\it mental representation} of the state of affairs in human society in general or 
in a social group of people. In particular, such a $\psi$ can be the mental representation of a real physical phenomenon. However, even 
in this case $\psi$ is not identified with the corresponding physical state. (By using the terminology invented 
by H. Atmanspacher and H. Primas, see, e.g.,
\cite{ATM}, we can consider the physical state as an ontic state and its mental image as an epistemic state.) This interpretation of representation 
of a state of the world by a pure quantum state matches well with the information interpretation of quantum mechanics (due to 
Zeilinger and Brukner).  Roughly speaking this $\psi$-function is not in nature, but in heads of people. See Remark 1 for further discussion.}  
is represented by $\psi$ and, for some $k_0,$  $P_\psi \leq P_{k_0}^{(i)},$ then, for the quantum probability
distribution corresponding to this state, we have: 
$$
p_\psi(P_{k_0}^{(i)})= \rm{Tr} P_\psi P_{k_0}^{(i)} =1\;  \rm{and, \: for}\; k\not=k_0,\;
p_\psi(P_{k}^{(i)})= \rm{Tr} P_\psi P_{k}^{(i)} =0. 
 $$ 
Thus, in this case, the event $P_{k_0}^{(i)}$ happens with the probability one and other events from  
information representation of the world by the $i$th agent have zero probability.

However, opposite to the classical case, in general $\psi$ need not belong to any concrete
subspace $H_{P_{k}^{(i)}}.$ Nevertheless, for any pure state $\psi$, there exists the minimal 
projector $Q_\psi^{(i)}$ of the form $\sum_{m} P_{j_m}^{(i)}$ such that $P_\psi \leq Q_\psi^{(i)}.$ 
Set $O_\psi^{(i)}=\{j: P_j^{(i)}\psi\not=0\}.$ Then
$Q_\psi^{(i)}= \sum_{j \in O_\psi^{(i)}} P_j^{(i)}.$  
The  projector $Q_\psi^{(i)}$ represents the $i$th agent's knowledge  about the $\psi$-world. 
We remark that $p_\psi(Q_\psi^{(i)})=1.$ 

Consider the system of projectors $\tilde{\cal P}^{(i)}$ consisting of sums of the projectors from ${\cal P}^{(i)}:$
\begin{equation}
\label{ha_TT99}
\tilde{\cal P}^{(i)} =\{P= \sum_m P^{(i)}_{j_m} \}.
\end{equation} 
Then 
\begin{equation}
\label{ha_TT}
Q_\psi^{(i)} = \min\{P \in \tilde{\cal P}^{(i)}: P_\psi \leq P\}. 
\end{equation}

{\bf Definition 1.} {\it 
For the $\psi$-state of the world and  the event $E,$ the $i$th agent knowns $E$ if}   
\begin{equation}
\label{ha}
Q_\psi^{(i)} \leq E.    
\end{equation}

\medskip

It is evident that if, for the state of the world $\psi,$ the $i$th agent knows $E,$ then $\psi \in H_E.$ In general
the latter does not imply that $E$ is known (for the state $\psi),$ see \cite{QI2014} for a discussion on definitions 
of knowing an event in the classical set-theoretic and quantum Hilbert space models.

\medskip

{\bf Remark 1.} For a single agent $i,$ ``quantumness'' is enconded in the possibility that the state of the world $\psi$ can be 
superposition of states belonging to different components of its information representation.   In the classical probabilistic 
framework knowing of an event $E$ means that, although an agent does not know precisely the state of the world $\omega,$ she/he 
knows precisely at least to which component $P_j$ this state belong. For quantum(-like) thinking agent,  a superposition state of the world does not 
give a possibility for ``precise orientation'' even in her/his information representation. 

\medskip

{\bf Example 1.} (Boeing MH17) For example, let us consider the case of 
the crush of Malaysian Boeing MH17 at Ukraine. As was pointed out in footnote 3, the state of the world $\psi$ represents the state of 
believes in society about possible sources of this crush. Suppose that there are only two possibilities: either the airplane was shut down 
by Keiv's military forces or by Donetsk's militants. For the illustrative purpose, it is sufficient to consider the two dimensional state 
space (although the real information state space related to the MH17-crush has a huge dimension depending on variety or political, economic, and 
military factors). Consider the basis $(e_K, e_D)$ representing the possibilities:  $e_K:$ ``Kiev is responsible'',  $e_D:$ 
``Donetsk is responsible''.  (We remark that in this model, if Kiev is reponsible than Donetsk is not and vise versa.)
In our model 
\begin{equation}
\label{ha_TT991}
\psi_{MH17}= c_1 e_K + c_2 e_D,
\end{equation}
where $c_1$ and $c_2$ complex probabilistic amplitudes for 
Kiev and Donetsk responsibilities, respectively. An agent tries to get know the truth about the MH17 crush by asking experts (say in terrorism).
\footnote{The first point is related to the discussion in footnote 3. The $\psi_{MH17}$ is not the actual physical state! The real physical state
of affairs can be (mentally) identified either with $e_K$ or with $e_D;$ the ontic state by the Atmanspacher-Primas terminology. However,
one has be careful in putting too much weight to the ontic state. It might happen that it would be never known.} She/he asked about 
their opinions; so the single question-observable
is in the use: ``Who is responsible?'' In the quantum model this agent operates with the spectral family ${\cal P}=\{P_1, P_2\},$ where
$P_1=P_{e_K}, P_2=P_{e_D}.$ If both amplitudes in (\ref{ha_TT991}) are nonzero (and in the present situation for July 24, 2014, it can 
be assumed that $c_1=c_2=1/\sqrt{2} ),$ then, for this state of the world, neither the event $E_K$ represented by $P_1$  nor
the event $E_D$ represented by $P_2$ is known (to be true) for this agent. In the classical model the state of the world $\omega$ has to
belong either to the element $P_1$ of the information partition or to the element $P_2.$ Thus one (and only one) of the events $E_K$ and
$E_D$ has to be known.

\bigskip

We now define the {\it knowledge operator} $K_i$ which applied to any event $E,$ yields the event 
``$i$th agent knows that $E.$'' 
 
\medskip

{\bf Definition 2.} {\it $K_i E= P_{H_{K_i E}},$ where $H_{K_i E}= \{\phi:  Q_{\phi/\Vert \phi \Vert}^{(i)} \leq E\}.$}     

\medskip

See \cite{QI2014} for the proof of the following proposition:

{\bf Proposition 1.} {\it For any event $E,$ the set $H_{K_i E}$ is a linear subspace of $H.$} 

Thus definition 2 is consistent.   The operator $K_i$ has the properties similar to the properties of the classical knowledge operator:

\medskip

{\bf Proposition 2.} {\it For any event $E,$}
\begin{equation}
\label{MMMOP}
{\cal K}1: \;\;\;K_i E \leq E.
\end{equation}

{\bf Proof.} Take nonzero $\phi \in H_{K_i E}.$ Then $Q_{\phi/\Vert \phi \Vert}^{(i)} \leq E$ and, hence, 
$$
H_{Q_{\phi/\Vert \phi \Vert}^{(i)}} \subset H_E.
$$
This implies that $\phi \in H_E$ and that $H_{K_i E} \subset H_E.$ 

\medskip
We also remark that trivially 

\begin{equation}
\label{MMMOP_0}
{\cal K}2: \;\;\; I \leq K_i I ,
\end{equation}
in fact,  
$$
I= K_i I.
$$

\medskip

{\bf Proposition 3.} {\it For any pair of events $E, F,$}
\begin{equation}
\label{MMMOP1}
E \leq F\; \mbox{implies}\;  K_i E \leq K_i F.
\end{equation}

{\bf Proof.} Take nonzero $\phi \in H_{K_i E}.$ Then $Q_{\phi/\Vert \phi \Vert}^{(i)} \leq E \leq F.$ Thus
$\phi \in K_i F.$

\medskip

{\bf Proposition 4.} {\it For any event pair of events $E, F,$}
\begin{equation}
\label{MMMOP2}
{\cal K}3: \;\;\; K_i E \wedge K_i F = K_i E\wedge F.
\end{equation}
{\bf Proof.} a). Take nonzero $\phi \in H_{K_i E} \cap H_{K_i F}.$ Then  $Q_{\phi/\Vert \phi \Vert}^{(i)} \leq E$ and 
$Q_{\phi/\Vert \phi \Vert}^{(i)} \leq F.$ Hence, $Q_{\phi/\Vert \phi \Vert}^{(i)} \leq E\wedge F$ and $\phi \in H_{K_i E\wedge F}.$
Therefore $K_i E \wedge K_i F \leq K_i E\wedge F.$

b). Take nonzero $\phi \in H_{K_i E \wedge F}.$ Then  $Q_{\phi/\Vert \phi \Vert}^{(i)} \leq E \wedge F$ and, hence, 
$Q_{\phi/\Vert \phi \Vert}^{(i)} \leq E$ and $Q_{\phi/\Vert \phi \Vert}^{(i)} \leq F.$ Therefore $\phi \in H_{K_iE} \cap H_{K_iE}=
H_{K_i E \wedge K_i F}$ and $K_i E\wedge F \leq K_i E \wedge K_i F .$

\medskip
{\bf Proposition 5.} {\it For any event $E,$}
\begin{equation}
\label{MMMOP3}
K_i E =\sum_{P_j^{(i)} \leq E} P_j^{(i)}.
\end{equation}

{\bf Proof.} a). First we show that $K_i E \leq \sum_{P_j^{(i)} \leq E} P_j^{(i)}.$ Take nonzero $\phi \in H_{K_i E}.$ Then $Q_{\phi/\Vert \phi \Vert}^{(i)} \leq E$ 
and $\phi= \sum_{j \in O_{\phi/\Vert \phi \Vert}^{(i)}} P_j^{(i)} \phi.$ Since $\sum_{j \in O_{\phi/\Vert \phi \Vert}^{(i)}} P_j^{(i)} \leq E,$
then for any $j \in O_{\phi/\Vert \phi \Vert}^{(i)},  P_j^{(i)} \leq E.$ Therefore 
$\phi= \sum_{P_j^{(i)} \leq E} P_j^{(i)} \phi.$

b). Now we show that $\sum_{P_j^{(i)} \leq E} P_j^{(i)} \leq K_i E.$ Let $\phi = \sum_{P_j^{(i)} \leq E} P_j^{(i)} \phi.$
Then  $Q_{\phi/\Vert \phi \Vert}^{(i)} \leq \sum_{P_j^{(i)} \leq E} P_j^{(i)} \leq E.$

\medskip

We also remark that  
\begin{equation}
\label{MMMOPi}
E =\sum  P_{j_k}^{(i)} \; \mbox{implies}\; K_i E = E.
\end{equation}
This immediately implies that 
\begin{equation}
\label{MMMOPia}
K_i E = K_i K_i E
\end{equation}
and, in particular, we obtain the following result (important for comparison
with the classical operator approach to definition of common knowledge):

{\bf Proposition 6.} {\it For any event $E,$}
\begin{equation}
\label{MMMOPi1}
{\cal K}4:  \;\;\; K_i E \leq  K_i K_i E.
\end{equation}

\medskip

Finally, we have:

{\bf Proposition 7.} {\it For any event $E,$}
\begin{equation}
\label{MMMOPi2}
(I- K_i E)  =  K_i(I- K_i E).
\end{equation}

{\bf Proof.} Take for simplicity that $K_iE= \sum_{j=1}^m P_{j}^{(i)},$ see (\ref{MMMOP3}). Then 
$I- K_i E = \sum_{j >m} P_{j}^{(i)}.$  By using (\ref{MMMOPi}) we obtain that 
$K_i(I- K_i E) =(I- K_i E).$

\medskip

In particular, we obtained that  
\begin{equation}
\label{MMMOPi3}
{\cal K}5: \;\;\; (I- K_i E)  \leq K_i(I- K_i E).
\end{equation}

\medskip

The  classical analogs of ${\cal K}1-{\cal K}5$ form the axiomatic base of the 
operator approach to common knowledge \cite{Brandenburger}. (Therefore we were so detailed in the presentation 
of ${\cal K}1-{\cal K}5;$ in particular, this aim, to match closer with the classical case, explains 
the above transitions from statements in the form of equalities, which are definitely stronger, to statements 
in the form of inequalities.) We also remark that in the classical approach to the knowledge operator the classical 
analog of the system 
${\cal K}1-{\cal K}5$ corresponds to the modal system $S5$ and 
of the system ${\cal K}1-{\cal K}4$  to the modal system $S4,$ see \cite{Kripke}. To analyze our quantum system ${\cal K}1-{\cal K}5$
from the viewpoint of its logical structure is an interesting and nontrivial problem. 

\medskip

{\bf Remark 2.} (Quantum truth?) This is a good place to discuss   
the truth content of quantum logic (which is formally represented as orthocomplemented closed subspace lattice of complex Hilbert space).
There are two opposite viewpoints on the truth content of quantum logic, see \cite{Garola}, \cite{Garola1} for the detailed discussion. From one viewpoint,
quantum logic carries not only the novel formal representation of knowledge about a new class of physical phenomena, but also assigns
to statements about these phenomena (at least to some of them) a special truth value, ``nonclassical truth''. Another viewpoint is that
one can proceed even in the  quantum case with the classical notion of truth as correspondence, which was 
explicated rigorously by Tarski's semantic theory, see    \cite{Garola}, \cite{Garola1}. 
The same problem states even more urgently in applications of the quantum formalism
in cognitive science and psychology: {\it Does quantum logic express new (nonclassical) truth assignment to propositions?}
Opposite to Garola et al.   \cite{Garola}, \cite{Garola1}, the authors of this paper consider quantum formalism as expressing
the new type of truth assignment, cf. \cite{Andrew}. However, the problem is extremely complex and it might happen that our position 
is wrong and the position of Garola \cite{Garola}, see also  Garola and Sozzo  \cite{Garola1}, is right. 
However, nowadays our approach is more common in discussions on the logical structure of quantum mechanics. 
It is usual in literature, e.g., \cite{KHR_CONT} to mention 
the use of different geometries, or probability theories, to uphold the thesis that also 
different logics could be needed in different physical theories.

{\bf Remark 3.} (Accessibility of quantum truth) The structures discovered in this paper are the formalization of 
the specific notion of common knowledge. Thus they do not by themselves formalize a notion of 
truth, but of a specific access to truth. Therefore, although the problem of whether the ``quantum truth'' can be reduced 
to the ``classical truth'' discussed in Remark 2 is important for clarification of quantum knowldege theory, it has no direct
relation to the subject of this paper.

\medskip

{\bf Definition 3.}
{\it Agent $i$'s possibility-projector ${\cal H}^{(i)}_\psi$ at the state of the world $\psi$ is defined as}
$$
{\cal H}^{(i)}_\psi = \bigwedge_{\{\psi \in K_i(E)\}} E. 
$$

It is easy to see that 
\begin{equation}
\label{MMMOPi4}
{\cal H}^{(i)}_\psi = Q^{(i)}_\psi. 
\end{equation}
It is interesting to point out that the collection of  $i$-agent's possibility-projectors (for all possible state) does 
not coincide with her spectral family and 
that different projectors are not mutually orthogonal. The latter is the crucial difference from the classical case.
In the latter any ``knowledge-map'' $K_i$ defined on the subsets of the set of 
states of the world, denoted as $\Omega,$ and satisfying axioms  ${\cal K}1-{\cal K}5$ generates possibility sets
giving disjoint partition of $\Omega.$ 
 
\medskip

Then, as in the classical case, we define:
$$
M_0 E = E, M_1 E= K_1 E \wedge...\wedge K_N E, ..., M_{n+1} E = K_1 M_n E \wedge ...\wedge K_N M_n E,...
$$
As usual, $M_1 E$  is the event ``all agents know that   $E$'' and so on.
We can rewrite this definition by using subspaces, instead of projectors:
$$
H_{M_1 E}= H_{K_1 E} \cap...\cap H_{K_N E},..., H_{M_{n+1} E}  = H_{K_1 M_n E} \cap ...\cap H_{K_N M_n E},...
$$
Now we define the {\it ``common knowledge''} operator, as mutual knowledge of all finite degrees:
$$
\kappa E = \wedge_{n=0}^\infty M_n E.
$$
Based on such quantum(-like) formalization of common knowledge, the validity of the Aumann theorem was analyzed 
in \cite{QI2014}.

\section{Possible generalization to multi-question information representations}

We considered a very special model of knowledge and common knowledge in which information representation of each agent $i$ 
is based on a {\it single question-operator} $A^{(i)}.$ Of course, it is natural to consider a more general model in which 
the $i$th agent can create his information representation based on the state of the world $\psi$ by using a few question-observables,
$A_k^{(i)}, k=1,...,M.$
First of all consider the case of compatible   observables, i.e., $[A_k^{(i)}, A_s^{(i)}]=0.$ Already in this case generalization of our model
is nontrivial and non-unique. 

First we recall how joint measurement of compatible observables is treated in quantum mechanics, starting with von Neumann\cite{VN}. Consider 
the case $M=2$ and omit the agent index $i.$ Thus the information representation is based on two question-observables which are 
mathematically represented by commuting operators $A_1$ and $A_2.$ There exists a Hermitian operator $R$ such that both operators 
can be represented as functions of $R: A_1= f_1(R), A_2=f_2(R).$ Then the joint measurement of these operators is reduced to measurement 
of the observable represented by $R$ and, for its value $r,$ the values $f_1(r)$ and   $f_2(r)$ are assigned to compatible question-observables.
Introduction of such a `` joint measurement operator'' $R$ completely washes out the individual spectral families of $A_k, k=1,2,$ which played 
the crucial role in the definitions
of knowledge/common knowledge. Suppose that the operator $R$ has the spectral decomposition
$$
R=\sum_j r_j P_j. 
$$
Then the corresponding knowledge model is simply based on the projectors ${\cal R}=(P_j).$ (Thus we get nothing new comparing with the previous
sections.) Consider the system of projectors $\tilde{\cal R}$ consisting of sums of the projectors from ${\cal R},$ 
see (\ref{ha_TT99}) (We work in the 
finite dimensional case, so all sums are finite.) For each state of the 
world $\psi,$  we introduce the projector      
\begin{equation}
\label{ha_TTkr99}
Q_{\psi} = \min\{P \in \tilde{\cal R}: P_\psi \leq P\}. 
\end{equation}
For the $\psi$-state of the world and  the event $E,$ the agent knowns $E$ if    
\begin{equation}
\label{ha_TTkr}
Q_{\psi} \leq E. 
\end{equation}
We call this model of knowing the von Neumann model.  

Although the presented scheme of measurement is the standard for quantum mechanics, it is not self-evident that precisely  this scheme 
have to be used as the basis for the quantum(-like) knowledge model corresponding to an agent operating with a family of questions 
represented by commuting operators. We propose another scheme which seems to be more natural for the quantum modeling of cognition. 
The main objection to application of the standard (von Neumann) quantum mechanical scheme of measurement for compatible observables is that 
in general an agent has not reason to try to construct the single observable such that both compatible question-observables can be expressed 
as its functions. Even if this is always possible theoretically, practically this process may be complicated and time consuming.    
An agent can prefer to proceed in testing knowing of an event $E$ by using each question separately. Mathematically this scheme 
is described as  follows.

Consider the spectral families of the question-operators (again we restrict consideration to the case of two operators), 
${\cal P}_1=(P_{1j})$ and ${\cal P}_2=(P_{2j})$ (we remind that the upper index corresponding to the agent
was omitted). Consider the systems of projectors $\tilde{\cal P}_k, k=1,2,$ consisting of sums of the projectors from ${\cal P}_k:$
$\tilde{\cal P}_k =\{P= \sum_m P_{k j_m} \}.$ 

For each state of the world $\psi$ and $k=1,2,$ we introduce the projectors      
\begin{equation}
\label{ha_TTk}
Q_{k; \psi} = \min\{P \in \tilde{\cal P}_k: P_\psi \leq P\}. 
\end{equation}

{\bf Definition 1A.} {\it 
For the $\psi$-state of the world and  the event $E,$ the agent knowns $E$ if}    
\begin{equation}
\label{ha_TTkn}
\mbox{either}\;  Q_{1; \psi} \leq E \;  \mbox{or}\;  Q_{2; \psi} \leq E. 
\end{equation}
It is clear that such knowing of $E$ implies its 
``von Neumann knowing'' based on (\ref{ha_TTkr99}), (\ref{ha_TTkr}). However, the inverse is not true.

\medskip

{\bf Example 2.} The state space $H$ of an agent is four dimensional with the orthonormal basis
$(e_1,e_2, e_3, e_4),$ the projectors $P_{11}$ and $P_{12}$ project $H$ onto the subspaces with the bases $(e_1,e_2)$ and 
$(e_3, e_4)$ and the  projectors $P_{21}$ and $P_{22}$ project $H$ onto the subspaces with the bases $(e_1,e_4)$ and 
$(e_2, e_3).$ The spectral family of the operator $R$ is given by one dimensional projectors $P_j=P_{e_j}.$ Consider 
the event $E$ given by the projector onto the subspace with the basis  $(e_1,e_2, e_3).$ Take the state of the world
$\psi= (e_1+e_2+e_3)/\sqrt{3}.$ Then $Q_\psi = E$ and the agent operating in the von Neumann scheme, i.e., who 
spent efforts to prepare the question-observable representing both compatible questions-operators, knows $E.$ 
However, the agent who produces knowledge by using two question-observables separately does not know $E.$ For him,
$Q_{1; \psi}= P_{11} + P_{12}= I$ as well as $Q_{1; \psi}= P_{21} + P_{22}= I.$    

\medskip

One of the advantages of the ``either/or'' scheme is that it has the straightforward generalization 
to incompatible observables, the same definition, Definition 1A.

\medskip

{\bf Example 3.} The state space $H$ of an agent is two dimensional. Consider in it two  orthonormal bases
$(e_{11},e_{12})$ and $(e_{21},e_{22})$ such that $\langle e_{1j}\vert e_{2m}\rangle \not=0$ 
 and the one-dimensional projectors corresponding 
to these bases, $P_{kj}= P_{e_{kj}}.$ Here ${\cal P}_k=\{P_{k1}, P_{k2}\}$ and $\tilde{\cal P}_k=\{P_{k1}, P_{k2}, I\},
k=1,2.$ Consider the event $E_1=P_{11}.$ Then this agent knowns it (``through the question observable with 
the spectral family   ${\cal P}_1$''.)  Consider the event $E_1=P_{21}.$ Then this agent knowns it (``through the question observable with 
the spectral family   ${\cal P}_2$''). Since projectors, for different $k,$ do not commute, there is no the ``joint measurement possibility'' and the 
operator $R$ does not exists, so the knowing scheme based on on (\ref{ha_TTkr99}), (\ref{ha_TTkr}) cannot be applied at all.     

\medskip

However, theory of such generalized knowledge operators is really beyond the scope of this paper.

\section*{Acknowledgments} 

The authors would like to thank  
C. Garola, E. Rosinger, and A. Schumann for resent exchange of ideas about the logical structure of quantum propositions.

\end{document}